\documentclass[12pt]{article}

\usepackage{amsmath,amsfonts,amssymb,latexsym,graphicx,float}
\usepackage{float}

\numberwithin{equation}{section}
\def\be{\begin{equation}}
\def\ee{\end{equation}}
\def\bq{\begin{eqnarray}}
\def\eq{\end{eqnarray}}
\def\beq{\begin{eqnarray}}
\def\eeq{\end{eqnarray}}

\begin{document}

\title{{\Large {\textsc{Localizing branes with bifurcating bulks}}}}
\author{{\large {\textsc{Ignatios Antoniadis}$^{1,2}$\thanks{%
antoniad@lpthe.jussieu.fr}, \textsc{Spiros Cotsakis}$^{3,4}$\thanks{%
skot@aegean.gr}, \textsc{John Miritzis}$^{5}$\thanks{%
imyr@aegean.gr}}} \\
$^1$Laboratoire de Physique Th\'{e}orique et Hautes Energies - LPTHE\\
Sorbonne Universit\'{e}, CNRS 4 Place Jussieu, 75005 Paris, France\\
$^2$ Department of Mathematical Sciences, University of Liverpool\\
Liverpool L69 7ZL, United Kingdom\\
$^3$Institute of Gravitation and Cosmology, RUDN University\\
ul. Miklukho-Maklaya 6, Moscow 117198, Russia\\
$^4$Research Laboratory of Geometry,\\
Dynamical Systems and Cosmology, University of the Aegean\\
Karlovassi 83200, Samos, Greece\\
$^{5}$Department of Marine Sciences\\
University of the Aegean\\
University Hill, Mytilene 81100, Greece}
\date{September 2022, final version}
\maketitle
\newpage
\begin{abstract}
\noindent We study the problem of evolution of bulk 5-fluids having an
embedded braneworld with a flat, de Sitter, or anti-de Sitter geometry. We
introduce new variables to express the Einstein equations as a
dynamical system that depends on the equation of state parameter  $\gamma$ and  exponent $\lambda$.
For linear fluids (i.e., $\lambda=1$), our formulation leads to a partial decoupling of the equations and thus to an exact solution. We find that such a fluid develops a transcritical bifurcation around the value $\gamma=-1/2$, and study how this behaviour  affects to stability of the solutions.
For nonlinear fluids, the situation is more diverse. We find an overall attractor at $\lambda=1/2$ and draw enough phase portraits to exhibit in detail the overall dynamics. We show that the value $\lambda =3/2$ is structurally unstable and typical for other forms of $\lambda$. Consequently, we observe a noticeable dependence of the qualitative behaviour of the solutions  on different `polytropic' forms of the fluid bulk. In addition, we prove the existence of a Dulac function for nonlinear fluids, signifying the impossibility of closed orbits in certain subsets of the phase space.
We also provide ample numerical evidence of gravity localizing solutions  on the brane which satisfy all energy conditions.

\end{abstract}
\newpage
\tableofcontents

\newpage

\section{Introduction}

It has been known for some time that singularity-free solutions are possible for the scale factor and thermodynamical quantities describing a 4-dimensional braneworld embedded in a 5-dimensional space with a fluid
analogue depending on the extra spatial coordinate and with a specific equation of
state \cite{add}-\cite{ack21a}. The problem is to find the
most general circumstances that allow such solutions with the properties of
satisfying the energy conditions and localising gravity on the brane.

In a series of works, we have been able to find special families of asymptotic
solutions that satisfy all the above-mentioned properties (cf. \cite{ackr1,ackr2}
and refs. therein). Although this search constitutes a viable approach to
the cosmological constant problem through the mechanism of self-tuning, the
search for the simplest solutions with the desired properties does not
reveal the structure of the whole space of interesting solutions.

In this paper, we assume that the bulk is non-compact and we provide a detailed study of  half of the space. In analogy with Ref.  \cite{rs2}, we look at the support of curved branes that localize gravity and satisfy the energy conditions. We use a bulk fluid-analog to make a model-independent analysis without specific field representation.

This setup leads us to consider the general problem of a 4-braneworld embedded in
a five-dimensional bulk space filled with a linear or nonlinear fluid from a
more qualitative, dynamical viewpoint. This provides us with an insight into the global
geometry of the orbits and the dynamics in the phase space of the problem. In addition, a more careful definition of the meaning of singularity-free solutions allows us to look into the problem from a more precise point of view.

The main difference of the present approach with earlier analyses is that
although previously our solutions were obtained as functions of the fifth
coordinate $Y$, in the present work we look for the global behaviour of
solutions as functions of the initial conditions. To achieve this goal, we
introduce new variables and a novel  formulation of the basic
brane-bulk equations.

These variables are analogues of the Hubble and the
density parameters $H, \Omega$ of relativistic cosmology, and are given as
functions of a suitable monotone reparametrization of $Y$. Also since the new variables contain the scale factor, its first derivative, as well as the density, they are able to provide a more precise picture of the possible singular solutions. The resulting  formulation
transforms the whole setup into a dynamical systems problem that can then be
studied using qualitative  methods.

The  reduction of the dynamics to the aforementioned  form
allows  interesting dynamical  properties to
be studied here for the first time in a brane-bulk phase space context. These include
the topological nature and bifurcations of equilibria, the phase portraits of the dynamics, the question of existence of closed orbits, as well as the dependence of the Planck-mass integral on initial
conditions.

The plan of this paper is as follows. In the next Section, we rewrite the
problem in terms of new  variables and arrive at dynamical
equations describing bulk fluids with an equation of state, and describe general features of the dynamics in Section 3.
In Sections 4, 5, we analyse the structure of dynamical solutions for
linear and nonlinear equations of state. In Section 6, we discuss the problem of localization of gravity on the brane, and we conclude with summarizing  our results in the last Section.

\section{Dimensionless formulation}

In this Section, we rewrite the basic dynamical equations in a new
dimensionless formulation for both the linear and the nonlinear fluid cases.

The five-dimensional Einstein equations on the bulk are given by,
\begin{equation}
G_{AB}=R_{AB}-\frac{1}{2}g_{AB}R=\kappa^{2}_{5}T_{AB},
\end{equation}
and we assume a bulk-filling fluid analogue with an energy-momentum tensor,
\begin{equation}
\label{T old}T_{AB}=(\rho+p)u_{A}u_{B}-p g_{AB},
\end{equation}
where the indices run from 1 to 5, the `pressure' $p$ and the `density' $\rho$ are functions only of the
fifth coordinate $Y$, and the fluid velocity vector field $u_{A}=\partial/\partial Y$
is parallel to the $Y$-dimension. We also choose units such that $\kappa_{5}=1$.
We consider below the evolution of this model for a brane-bulk metric given
by
\begin{equation}
\label{warpmetric}g_{5}=a^{2}(Y)g_{4}+dY^{2},
\end{equation}
where $a(Y)$ is a warp factor with $a(Y)>0$, while the brane metric $g_{4}$ is taken to be the
four-dimensional flat, de Sitter or anti-de Sitter standard metric.

With this setup,  the Einstein equations split into the conservation equation,
\begin{equation}
\label{syst2iii}\rho^{\prime}+4(p+\rho)\frac{a^{\prime}}{a}=0,
\end{equation}
the Raychaudhouri equation,
\begin{equation}
\label{syst2ii}\frac{a^{\prime\prime}}{a}=-\frac{1}{6}%
{(2p+\rho)},
\end{equation}
and the Friedmann equation,
\begin{equation}\label{syst2i}
6\frac{a^{\prime2}}{a^{2}}=\rho+\frac{6k}{a^{2}},
\end{equation}
which is a first integral of the other two when $a^{\prime}\neq0$. Here, the
constant $k$ is zero for a flat brane, $k=1$ for a de Sitter brane, and $k=-1$
for an anti de Sitter brane. For the brane-bulk problem, there is also the junction condition which in general describes a jump discontinuity in the first derivative of $a(Y)$ and takes the generic form,
\be \label{j1}
a'(0^+)-a'(0^-)=f(a(0),\rho(0)),
\ee
where the brane tension $f$ is a continuous, non-vanishing function of the initial values of $a,\rho$ (for examples of this in specific solutions, see Ref. \cite{ackr1,ackr2}).

Inspired by standard cosmology, we introduce the following `observational
parameters' related to the extra `bulk' dimension $Y$: The \emph{Hubble
scalar} $H$, measuring the rate of the expansion (a prime $^{\prime}$ means
$d/dY$),
\begin{equation}
H=\frac{a^{\prime}}{a}, \label{h}%
\end{equation}
the \emph{deceleration parameter} $q$, which measures the possible speeding up
or slowing down of the expansion in the $Y$ dimension,
\begin{equation}
q=-\frac{a^{\prime\prime}a}{a^{\prime2}}, \label{q}%
\end{equation}
and the \emph{density parameter} $\Omega$ describing the bulk matter density
effects,
\begin{equation}
\Omega=\frac{\rho}{3H^{2}}. \label{o}%
\end{equation}
While $\Omega,q$ are dimensionless, the Hubble scalar $H$ has dimensions
$[Y]^{-1}$.


Using these variables and dividing both sides by $3H^2$, the Friedmann equation (\ref{syst2i}) is,
\be\label{constr}
2-\Omega=\frac{2k}{H^{2}a^{2}},
\ee
and we conclude that evolution of the 5-dimensional models  with
\begin{itemize}
\item $\Omega>2$ corresponds to those having an AdS brane ($k=-1$)
\item $\Omega=2$ corresponds to those having a flat brane ($k=0$)
\item $\Omega<2$ corresponds to those having a dS brane ($k=+1$).
\end{itemize}
(By redefining $\lambda_{5}^{2} =\kappa_{5}^{2} /2$, we would have obtained  the usual trichotomy $\Omega\gtreqqless 1$ relations here, but this would have also changed various coefficients in the other field equations, so we prefer to leave it as above.)

We set $Y_{0}$ for some
arbitrarily chosen reference value of the bulk variable $Y$, and
$a_{0}=a(Y_{0})$. The evolution will be described not in terms of $Y$ but by a
new dimensionless bulk variable $\tau$ in the place of $Y$, with,
\begin{equation}
\label{dimless1}a=a_{0} e^{\tau}.
\end{equation}
Then we have,
\begin{equation}
a^{\prime}=a\,\frac{d\tau}{dY},
\end{equation}
and hence,
\begin{equation}\label{dimless2}
\frac{dY}{d\tau}=\frac{1}{H}.
\end{equation}
We shall assume that the brane lies at
$\tau=0$, or  $a=a_{0}$ (we may assume without loss of generality that $Y_0=0$). In this case, the junction condition (\ref{j1}) expressed in terms of the new variables $H,\Omega$ implies that,
\be \label{j2}
H(0^+)-H(0^-)=f(x(0),0),\quad x=H^2\Omega,
\ee
where the brane tension $f$ is a function of the variables $x,\tau$. The variable $x$ is regular from the definition (\ref{o}), and $\tau$ is regular from (\ref{dimless2}) because $H$ has only a finite discontinuity at $0$ as it follows from the junction condition (\ref{j1}).

Having a brane located at the length scale value
$a_{0} $, with $0<a<+\infty$, we consider two intervals, $\tau\in(-\infty,0)$
- the `left-side' evolution, and the `right-side' interval $\tau\in
(0,+\infty)$. The latter is equivalent to the left-side interval under the
transformation $\tau\rightarrow-\tau$, and so without loss of generality we
shall restrict our attention only to the $\tau$-range $(-\infty,0)$. All our
results involving the dimensionless variable $\tau$ can be transferred to the
right-side interval by taking $\tau\rightarrow-\tau$. This will be important later.

We now show that the dynamics of the system (\ref{syst2iii})-(\ref{syst2i})
can be equivalently described by a simpler dynamical system in terms of $H$ and the
dimensionless variables  $q,\tau $, and $\Omega$. From the above definitions for $H$ and $q$, we are led to the following
evolution equation for $H$, namely,
\begin{equation}
\frac{dH}{d\tau}=-(1+q)H.
\end{equation}
In the following we shall assume that fluids in the bulk satisfy linear or
nonlinear equations of state. Then the $q$-equation for a linear equation of
state (EoS),
\begin{equation}
p=\gamma\rho, \label{linro}%
\end{equation}
is found to be,
\begin{equation}
q=\left(  \frac{1}{2}+\gamma\right)  \Omega, \label{qlin}%
\end{equation}
while for the nonlinear equation of state,
\begin{equation}
p=\gamma\rho^{\lambda}, \label{nonlinro}%
\end{equation}
for some parameter $\lambda$, we have,
\begin{equation}
q=\left(  \frac{1}{2}+\gamma\rho^{\lambda-1}\right)  \Omega. \label{qnonlin}%
\end{equation}
We note here the usual fluid parameter $\gamma$ will be constrained later using the energy conditions. On the other hand, the parameter $\lambda$, the ratio $c_P/c_V$ of the specific heats of the bulk fluid under constant pressure and volume, is  commonly taken to satisfy $\lambda>1$ in other contexts, most notable in standard stellar structure theory (cf. e.g., \cite{cha}, chap. IV). Although we generally put no constraint on it, we shall find that in the present problem there is a clear preference for the `polytropic' values $\lambda=1+1/n$, for integer $n$, valid for the entire bulk. Such polytropic changes provide ample differences as compared to the `collisionless' case $n=\infty$, for small values of $n$.

The evolution equation for the Hubble scalar for the linear EoS case
(\ref{linro}), becomes,
\begin{equation}
\frac{dH}{d\tau}=-\left(  1+\left(  \gamma+\frac{1}{2}\right)  \Omega\right)
H, \label{hlin}%
\end{equation}
whereas the nonlinear equation of state case (\ref{nonlinro}), gives the
Hubble evolution equation in the form,
\begin{equation}
\frac{dH}{d\tau}=-H-\frac{\Omega H}{2}-3^{\lambda-1}\gamma H^{2\lambda
-1}\Omega^{\lambda}. \label{hnonlin}%
\end{equation}
Let us lastly consider the continuity equation (\ref{syst2iii}). This
equation, assuming that $H\neq0$, in the linear-fluid case becomes,
\begin{equation}
\frac{d\Omega}{d\tau}=2(q-2\gamma-1)\Omega. \label{olin}%
\end{equation}
On the other hand, in the nonlinear-fluid case, assuming again that $H\neq0$,
and using Eq. (\ref{hnonlin}), we find the following evolution equation for
$\Omega$, namely,
\begin{equation}
\frac{d\Omega}{d\tau}=-2\Omega+\Omega^{2}+2\gamma3^{\lambda-1}H^{2\lambda
-2}\Omega^{\lambda+1}-4\gamma3^{\lambda-1}H^{2\lambda-2}\Omega^{\lambda}.
\label{ononlin}%
\end{equation}

Summarizing, in our new  formulation of the bulk-brane problem, the basic dynamical systems
are given by the Friedman constraint Eq. (\ref{constr}), together with evolution equations in the following forms.

\textbf{Case A: Nonlinear EoS, $p=\gamma\rho^{\lambda}$.} In this case, we
have a 2-dimensional dynamical system, namely,%
\begin{align}
\frac{dH}{d\tau}  &  =-H-\frac{\Omega H}{2}-3^{\lambda-1}\gamma H^{2\lambda
-1}\Omega^{\lambda},\label{nls1}\\
\frac{d\Omega}{d\tau}  &  =-2\Omega+\Omega^{2}+2\gamma3^{\lambda-1}%
H^{2\lambda-2}\Omega^{\lambda+1}-4\gamma3^{\lambda-1}H^{2\lambda-2}%
\Omega^{\lambda}. \label{nls2}%
\end{align}

\textbf{Case B: Linear EoS, $p=\gamma\rho$.} This is the special case with
$\lambda=1$. We have the $H$ equation,
\begin{equation}
\frac{dH}{d\tau}=-\left(  1+\left(  \gamma+\frac{1}{2}\right)  \Omega\right)
H, \label{ls1}%
\end{equation}
and a single, decoupled evolution equation for $\Omega$, namely,
\begin{equation}
\frac{d\Omega}{d\tau}=2\left(  \left(  \gamma+\frac{1}{2}\right)
\Omega-2\gamma-1\right)  \Omega. \label{ls2}%
\end{equation}

\section{General properties}
The 5-dimensional fluid solutions are then given in terms of the $(H,\Omega)$ variables which satisfy these evolution equations together with the Friedman constraint Eq. (\ref{constr}). For their physical interpretation,  it is helpful to use a new classification in terms of $H, \Omega$ and $k$. We shall say that the bulk fluid is:
\begin{enumerate}
\item \emph{a dS (resp. an AdS) fluid}, when $k=+1$ (resp. $k=-1$)
\item a \emph{flat fluid}, when $k=0$.
\item \emph{Static}, when $H=0$ (it is necessarily flat in this case).
\item \emph{Expanding (resp. contracting)}, when $H>0\, (<0)$. (This means that the fluid is moving away (resp. towards) the brane for positive $\tau$)
\end{enumerate}
As we have noted after Eq. (\ref{constr}), the cases $\Omega <2, =2, >2$ correspond to a dS, a flat, or a AdS fluid respectively, while in the case $\Omega=0$, the bulk is empty, and the constraint equation (\ref{constr}) necessarily implies $k=+1$ for consistency, hence,  $a(\tau)=\tau+C$ in this case. We shall refer to the $\Omega=0$ case as an \emph{empty dS bulk}.

We shall also invariably refer to any given phase point  $(H,\Omega)$ as a `state', for example the point $(0,0)$ describes the  state of as static, empty dS bulk, while the $(0,2)$ state is a static, flat fluid. Dynamical (non-static) states require $H\neq 0$, and these are described as non-trivial orbits in the  $(H,\Omega)$ phase space. Further classification tags for each one of these models appear in the next Sections and depend on the ranges and values of the two fluid parameters $\gamma,\lambda$ that appear in the evolution equations.

Some general remarks about the dynamical system (\ref{nls1})-(\ref{nls2}), and its
special case (\ref{ls1})-(\ref{ls2}) are in order:
\begin{itemize}
\item Since the dynamical systems studied in this paper are two-dimensional (with a constraint), our search is for bifurcations, oscillations, or limit cycles, but no chaotic behaviour, strange attractors,  or more complex phenomena can be present here.
\item Equation (\ref{nls2}) implies that the set $\Omega=0$ is invariant under
the flow of the dynamical system, i.e. $\Omega=0$ is a solution of the system.
Since no trajectories of the dynamical system can cross, we conclude that if
initially the state of the system is on the line $\Omega=0$ (that is if we start with an `empty dS bulk'), it will remain on
this line for ever. Therefore, if initially $\Omega$ is positive, it remains
positive for ever.  We emphasize that this result holds for all
$\lambda\geq0$.

\item If $\lambda\geq1$, equation (\ref{nls1}) implies that the set $H=0$ is
also invariant under the flow of the dynamical system. Therefore, for
$\lambda\geq1$, assuming that initially $H>0$, then $H\left(  \tau\right)  >0$
for all $\tau\geq0$. We conclude that any trajectory starting at the first
quadrant $H\geq0,$ $\Omega\geq0,$ cannot cross the axes and therefore, cannot
escape out of this quadrant. For instance, expanding AdS  fluids, and expanding empty dS bulks remain always expanding, and static AdS  fluids always remain static.

\item For the linear EoS case, it is important that the $\Omega$-equation
(\ref{ls2}) decouples, that is, it does not contain the $H$. This decoupling is due to our choice of new variables (cf. Eq. (\ref{dimless1})). So now the $\Omega$-equation can be
treated separately as a logistic-type equation, and this simplifies the
analysis considerably. In this case, solving the $\Omega$-equation and substituting in the
$H$-equation, we have a full solution of the system. This feature is absent
from the nonlinear EoS fluid equations, which comprise a truly coupled $2D$
system.  This is studied more fully below.

\item The necessity of satisfying the energy conditions (cf. \cite{ackr1}) leads in general to restrictions on the $\gamma$ range. For a fluid with a linear EoS the intersection of the requirements that follow from the weak, strong, or null energy conditions lead to the typical range $\gamma\in [-1,1]$. We shall assume this restriction as a minimum requirement for our acceptance of a solution property.
\end{itemize}
Below, with an slight abuse of language, we shall use the term \emph{linear (nonlinear) fluid} when the respective EoS is linear (nonlinear).

\section{Linear fluids and their bifurcations}
This Section provides a study of the behaviour of bulk fluids with the linear EoS given by Eq. (\ref{linro}) and  described by the nonlinear system (\ref{ls1}), (\ref{ls2}), (\ref{constr}).

This system can be solved exactly and the asymptotic properties of the solutions displayed graphically. Equation (\ref{ls2}) has the form
\begin{equation}
\frac{d\Omega}{d\tau}=A\Omega^{2}+B\Omega,\label{function}%
\end{equation}
with $A=2(\gamma+1/2),~B=-2A$.
The $\Omega$-solution from Eq. (\ref{function}) with initial condition
$\Omega\left(  0\right)  =\Omega_{0}$ is given by%
\begin{equation}
\Omega\left(  \tau\right)  =\frac{B/A}{\left(  \frac{B}{A\Omega_{0}}+1\right)
e^{-B\tau}-1}=\frac{2\Omega_{0}}{\Omega_{0}+\left(  2-\Omega_{0}\right)
e^{2\left(  2\gamma+1\right)  \tau}},\label{omega}%
\end{equation}
The resulting  $H$-solutions are found by substituting in (\ref{ls1}) (which is a linear differential equation), yielding,
\begin{equation}
H\left(  \tau\right)  =\frac{H_{0}e^{-\tau}\sqrt{\Omega_{0}e^{-2\left(
2\gamma+1\right)  \tau}-\Omega_{0}+2}}{\sqrt{2}}, \label{hubble}%
\end{equation}
where $H_{0}=H\left(  0\right) $.

From these solutions it follows that dS fluids ($2-\Omega_0>0$) are  real solutions for all values of the fluid parameter $\gamma$ and signs of $\tau$. However, the AdS solutions ($2-\Omega_0<0$) become complex when $\gamma>-1/2, \tau>0$ or when  $\gamma<-1/2, \tau<0$, since in these cases the exponential in the right-hand-side of Eq. (\ref{hubble}) decays. Hence, AdS fluids  are real only when $\gamma<-1/2, \tau>0$, or when $\gamma>-1/2, \tau<0$, and we shall consider AdS solutions only in these ranges.

For dS fluids, we can see that  $\Omega\left(\tau\right)$ approaches zero when $\gamma>-1/2, \tau>0$ or  $\gamma<-1/2, \tau<0$, and approaches two when $\gamma<-1/2, \tau>0$, or when $\gamma>-1/2, \tau<0$. All AdS fluids  develop $\Omega$ blow up singularities for $\gamma<-1/2, \tau>0$, or  $\gamma>-1/2, \tau<0$.

\begin{figure}[tbh]
\begin{center}
\includegraphics[scale=1]{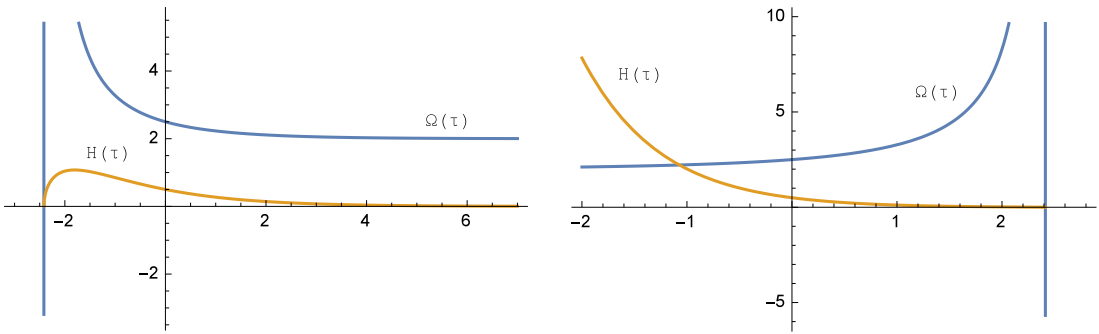}
\end{center}
\vspace*{-20mm}
\caption{Solutions (\ref{omega}) for $\gamma<-1/2$ and $\gamma>-1/2$ when
$\Omega_{0}>2$.}%
\label{oh1}%
\end{figure}

\begin{figure}[tbh]
\begin{center}
\includegraphics[scale=1]{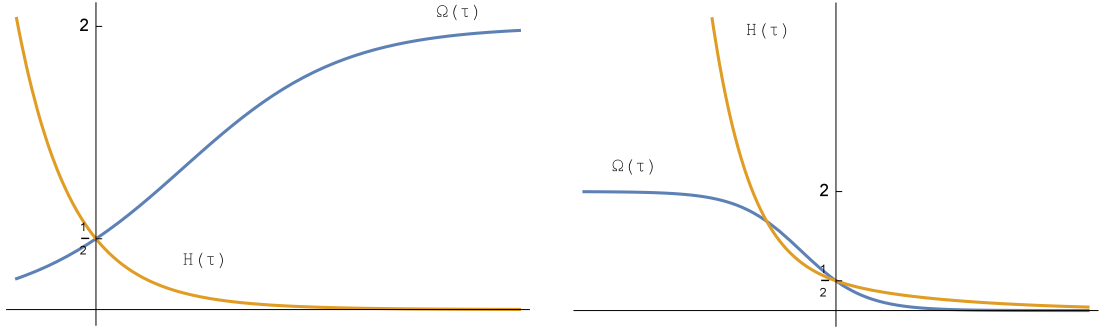}
\end{center}
\vspace*{-20mm}
\caption{Solutions for  $\gamma=-2/3$ and $\gamma=1/2$. In the first case the solution (\ref{omega})
increases following the usual logistic curve and quickly approaches the
constant value $2$.}%
\label{oh}%
\end{figure}

On the other hand, to disclose the asymptotic nature of the $H$-solutions we can look at the monotonicity properties of the function $H(\tau)$ and its possible dependence on different $\gamma$ ranges.
The results  show a further sensitive dependence on the $\gamma$ parameter around its
$\gamma=-1$ value\footnote{For $\gamma<-1$, the term $-2\left(  2\gamma+1\right)  $ is
positive (and $>2$), therefore, the term $e^{-\tau}$ dominates over
$\sqrt{e^{-2\left(  2\gamma+1\right)  \tau}}$ for $\tau<0$, while the opposite
happens for $\tau>0$. Thus, the solution $H\left(\tau\right)$ is
decreasing for $\tau<0$ and is increasing for $\tau>0$. As discussed already, solutions in this range of $\gamma$ are not acceptable as they do not satisfy the energy conditions, and so we shall not consider them further.}. For $\gamma>-1$,  $H\left(  \tau\right)  $ always decreases and
approaches zero. At the critical value $\gamma=-1$ the solution $H\left(
\tau\right)$ decreases and approaches the constant
$H_{0}\sqrt{1-\Omega_{0}/2},$ provided that $\Omega_{0}<2$. We note that the behaviour of the solutions
(\ref{omega}) and (\ref{hubble}) is insensitive on the initial value $H_{0}$. The asymptotic behaviours of the $H,\Omega$ solution is shown in Figures \ref{oh1}, \ref{oh}.

The nature of the  solutions is further revealed by studying the stability of the equilibrium solutions. This is effected by formally setting $X=(H,\Omega)$, and think
of the system (\ref{ls1})-(\ref{ls2}) as one of the form,
\begin{equation}
\frac{dX}{d\tau}=F(X,\gamma),\quad F=(F_{1},F_{2}),
\end{equation}
where $\gamma\in(-\infty,\infty)$, and with the $F_{i},i=1,2$, being the right-hand-sides
of Eqns. (\ref{ls1}) and (\ref{ls2})  respectively. With this notation, we now show that the solutions of the system undergo a transcritical bifurcation when $\gamma=-1/2 $ at the origin which is a
non-hyperbolic equilibrium. This means that bulk fluids exchange their stability when the EoS parameter $\gamma$ passes through $-1/2$.

\begin{figure}[tbh]
\begin{center}
\includegraphics[scale=0.7]{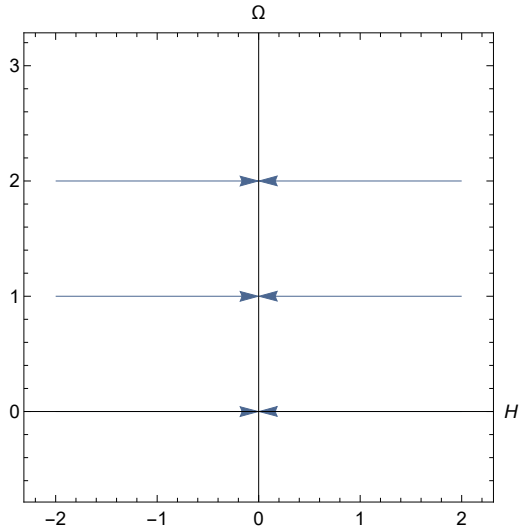}
\end{center}
\vspace*{-10mm}
\caption{Phase portrait of the system (\ref{ls1})-(\ref{ls2}) for
$\gamma=-1/2$. Every point on the $\Omega$ axis is an equilibrium.}%
\label{phase-linear-half}%
\end{figure}

Returning to Eq.  (\ref{ls2}), when $\gamma=-1/2$, the system is $\Omega^{\prime}=0,H^{\prime}=-H$, with
immediate solution,
\begin{equation}
\Omega\left(  \tau\right)  =\Omega_{0}=\mathrm{const.},\ \ H\left(
\tau\right)  =H_{0}e^{-\tau}. \label{gammaminus1}
\end{equation}
In this case,   every point on the $\Omega$ axis is a non-hyperbolic equilibrium and every other point on the phase plane approaches the corresponding point of the $\Omega$ axis as shown in Fig \ref{phase-linear-half}. Therefore in this case we have expanding universes with ever-decreasing rate and collapsing ones with an ever-increasing rate both approaching a static dS bulk of a constant density. As two of us have shown elsewhere (cf. Ref. \cite{ackr1}, Sect. 4.1.1), for this value of $\gamma$ the system satisfies all the energy conditions.

\begin{figure}[tbh]
\begin{center}
\includegraphics{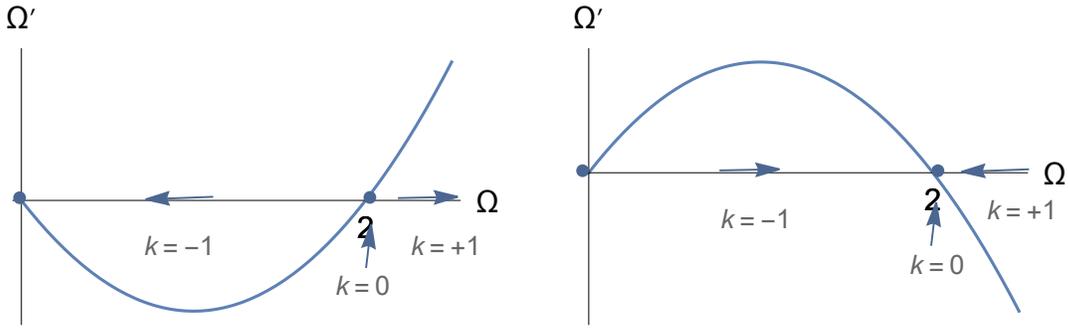}
\end{center}
\vspace*{-10mm}
\caption{The 1-dimensional phase space ($\Omega$-line) in the linear fluid
case. The Figure on the left (right) shows the case $\gamma>-1/2 $
($\gamma<-1/2$). The arrows correspond to evolution in positive $\tau$.}%
\label{onedim}%
\end{figure}

\begin{figure}[tbh]
\begin{center}
\includegraphics[scale=0.8]{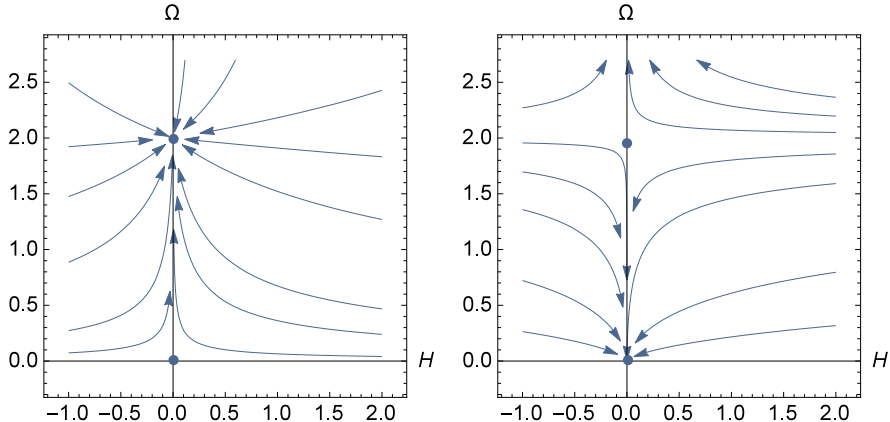}
\end{center}
\vspace*{-5mm}
\caption{Phase portrait of the system (\ref{ls1})-(\ref{ls2}) for
$\gamma<-1/2$ and $\gamma>-1/2$. The stable node at $(0,2) $ (left), exchanges
stability with the saddle $( 0,0) $ (right) as the $\gamma$ values change.}%
\label{phase_linear}%
\end{figure}

The case $\gamma\neq-1/2 $ is shown in Figures \ref{onedim}, \ref{phase_linear}. When $\gamma
\gtrless-1/2 $, we have $F_{2}^{\prime\prime}\gtrless0$, and so $F_{2}%
(\Omega)$ is a convex or a concave function respectively. In this case, there
are two equilibria, one at $\Omega=0$, and a second one at $\Omega=2$. When
$\gamma>-1/2 $, the equilibrium at the origin is stable while the one at
$\Omega=2$ is unstable, and they exchange their stability when $\gamma<-1/2$.
For initial conditions with $\Omega
_{0}<2$, that is for bulk models with a dS brane, and for the case
$\gamma>-1/2$ (the left diagram in Fig. \ref{onedim}), the solution $\Omega\left(
\tau\right)  $ decreases approaching zero, the `Milne state' for positive
$\tau$, whereas for $\gamma<-1/2$ (the right diagram in Fig. \ref{onedim}), the solution
(\ref{omega}) increases and approaches the constant value $2$ which
corresponds to a flat brane ($k=0$).
The situation is different if initial conditions with $\Omega_{0}>2$, that is
for bulk models with a AdS brane ($k=+1$) are considered. For $\gamma>-1/2$,
the solution $\Omega\left(  \tau\right)  $ increases without bound, whereas
for $\gamma<-1/2$ the solution $\Omega\left(  \tau\right)  $ decreases to the
flat state at $\Omega=2$. Therefore we have a transcritical bifurcation occurring at the parameter value
$\gamma=-1/2$, so that the two equilibria switch their stability without
disappearing after the bifurcation, see Figure  \ref{phase_linear} for the full
phase portrait of the system.

These results when translated to the bulk-brane language imply that the evolution of  bulk fluids with a linear equation of state depends of the $\gamma$ parameter and is organized around the two simplest equilibria, namely, the empty bulk and the flat fluid, which exchange their stability because of the transcritical bifurcation as the nature of the fluid changes (depending on $\gamma$). A typical bulk fluid evolves either towards or away from the equilibrium states `empty bulk' and `flat fluid' depending on whether  it has $\gamma\gtrless-1/2$ as shown in Fig. \ref{phase_linear}.


\begin{figure}[tbh]
\begin{center}
\includegraphics[scale=0.6]{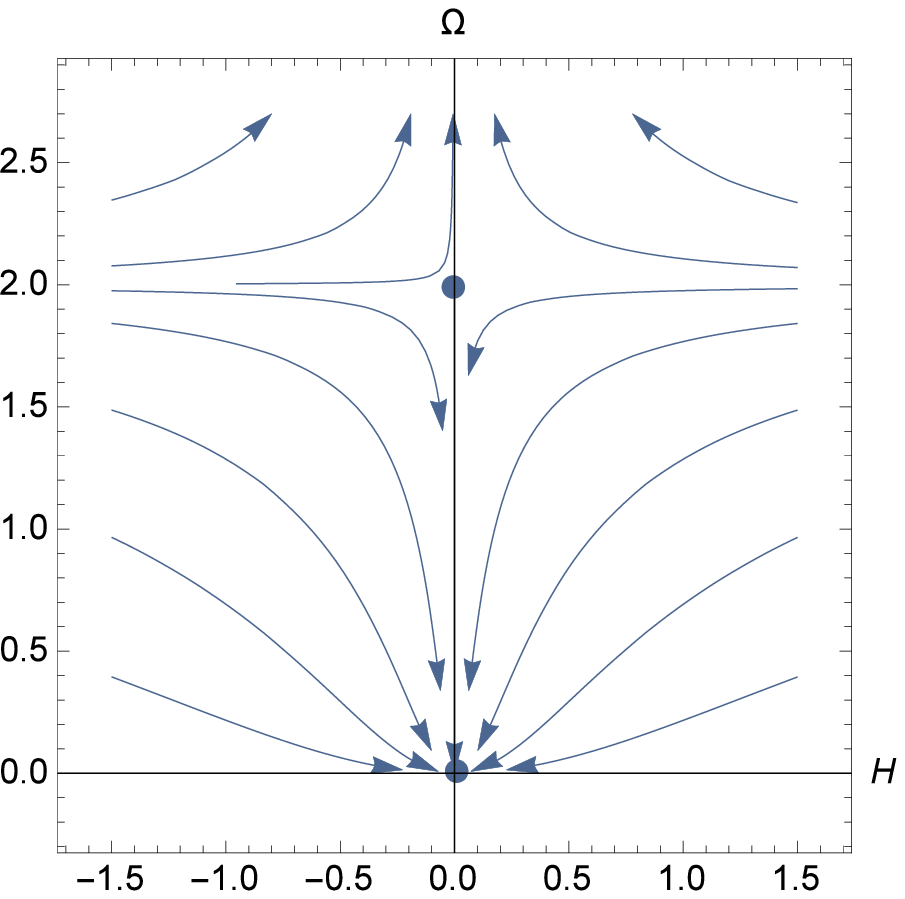}
\end{center}
\vspace*{-8mm}
\caption{Phase portrait of the system (\ref{nls1})-(\ref{nls2}) for arbitrary
$\lambda$ and $\gamma=0$. }%
\label{gammazero}%
\end{figure}

A last special case of importance is that of a `bulk dust'. The corresponding behaviour of a bulk fluid with a linear EoS is also shared in this case with all bulk fluids having a  nonlinear EoS (see next Section).  For $\gamma=0$, the system (\ref{nls1})-(\ref{nls2}) reduces to
\be
\begin{aligned}
\frac{dH}{d\tau} &  =-H-\frac{\Omega H}{2},\\
\frac{d\Omega}{d\tau} &  =-2\Omega+\Omega^{2},\label{gamma0}%
\end{aligned}
\ee
for all $\lambda$. The phase portrait of the system (\ref{gamma0}) indicates
that all solutions with initial values $\Omega_{0}<2$ (this corresponds to an  dS dust fluid) and $H_{0}$ arbitrary,
asymptotically approach the node $\left(  0,0\right)  $ (that is a static, empty dS bulk), see Figure
\ref{gammazero}. Hence, we find that dust-filled dS bulks rarefy to empty ones. This can be proved by noting that the formal solution of the
first of (\ref{gamma0}) can be written as
\be
H\left(  \tau\right)  =H\left(  0\right)  \exp\left(  -\int_{0}^{\tau}\left(
1+\Omega\left(  s\right)  /2\right)  ds\right)  ,
\ee
which goes to zero as $\tau\rightarrow\infty$. By the same formula, we can see again here (like in Fig. \ref{phase_linear}, right phase portrait)
that AdS trajectories starting above the line $\Omega_{0}=2$ approach the $\Omega$
axis, while $\Omega\left(  \tau\right)  $ diverges.

\section{Nonlinear fluids: Regularity and stability}

Let us now move to discuss properties of the nonlinear, two-dimensional system
(\ref{nls1})-(\ref{nls2}). In distinction to the linear case treated above,
this is a genuine, coupled, two-dimensional dynamical system and this results
in two important effects that we discuss below. We first discuss the nature of
the equilibria of the system and study the phase portrait. We then find a
suitable Dulac function for the dynamics of the nonlinear fluid-brane system
and show that there can be no closed (periodic) orbits for the system in certain parts of the phase space.

The nature of the $(H,\Omega)$-solutions of the system (\ref{nls1}%
)-(\ref{nls2}) is strongly dependent on the ranges of the $\lambda$-parameter
present in the fluid's nonlinear equation of state Eq. (\ref{nonlinro}), in
particular, on the three ranges, $\lambda<0$, $\lambda\in(0,1)$, and
$\lambda\geq1$.

When $\lambda<0$, there are no finite equilibria for the system (\ref{nls1}%
)-(\ref{nls2}). In this case, the dynamics is transferred to points at
infinity, a more complicated problem that we do not consider in this paper,
since it requires a deeper analysis of the `companion system' to
(\ref{nls1})-(\ref{nls2}), cf. \cite{go1,cb07}. Because of the presence of
denominators in the vector field that defines the system (\ref{nls1}%
)-(\ref{nls2}) when $\lambda<0$, an analysis of this case will help to further
clarify the question of the existence of stable singularity-free solutions of
the system. We only further note that the `dynamics at infinity' in this case
may be realized through the Poincar\'{e} sphere compactification as a boundary
dynamics in the framework of \emph{ambient cosmology}, an extension of brane
cosmology wherein the brane lies at the conformal infinity of the bulk
\cite{ac15}.

Next, for the case $\lambda>0$, we must distinguish between the two cases $\lambda\in(0,1)$ and $\lambda\geq 1$. For $\lambda\geq 1$,  there are always  the two $\gamma$-independent equilibria at $(0,0)$ and $(0,2)$. In addition, there are $\gamma$-dependent equilibria being generally complex:
\begin{equation}
\left(  H_{\ast},\Omega_{\ast}\right)  =\left(  \exp\left(  \frac
{-i\pi+\lambda\ln6+\ln\frac{\gamma}{6}}{2-2\lambda}\right)  ,2\right)
.\label{equilibria}%
\end{equation}
These equilibria are real provided $\lambda$ takes the values
\begin{equation}
\lambda(n)=1\pm\frac{1}{2n},\quad n=1,2,\dots.\label{lambda}%
\end{equation}
It is interesting that this case falls into the polytropic index form of the $\lambda$ exponent (cf. \cite{sc}, section 8).
For $\lambda$ taking the values $\lambda(n)=1-1/2n$, the equilibria
(\ref{equilibria}) all correspond to flat bulk fluids as they lay on the line
$\Omega=2$, and at the points where,
\begin{equation}
H_{\ast}=\left\{  -\frac{\gamma}{\sqrt{6}},\frac{\gamma^{2}}{\sqrt{6}}%
,-\frac{\gamma^{3}}{\sqrt{6}},\frac{\gamma^{4}}{\sqrt{6}},...\right\}
.\label{hequ}%
\end{equation}
For the remaining values of $\lambda$ in the case where $\lambda\in(0,1)$, there are no other finite equilibria, and the presence of denominators in the vector field, as in the case of $\lambda<0$, makes this case more complicated dynamically.

\begin{figure}[tbh]
\begin{center}
\includegraphics[scale=0.8]{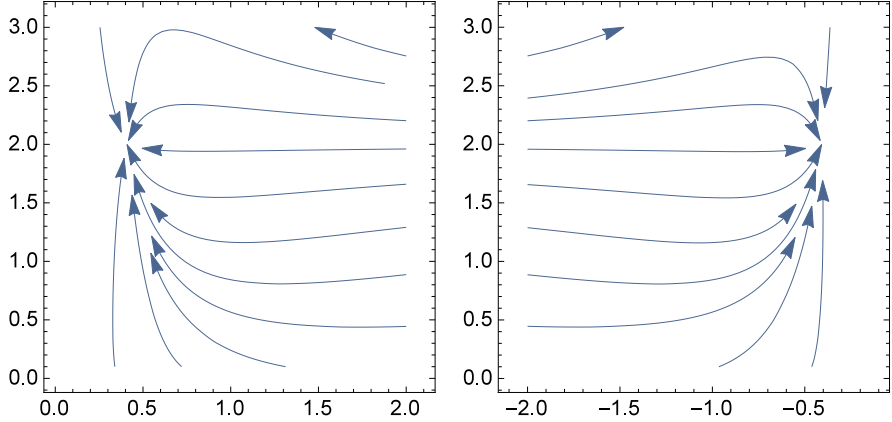}
\end{center}
\par
\vspace*{-10mm} \caption{Phase portrait of the system (\ref{nls1}%
)-(\ref{nls2}) for $\lambda=1/2$ and $\gamma=-1$ and $\gamma=1$. }%
\label{node1}%
\end{figure}

To construct a typical phase portrait for $\lambda=1-1/2n$, we
may use $\lambda(1)=1/2$. The only equilibrium of the system is $\left(
-\gamma/\sqrt{6},2\right)  $ as implied by (\ref{hequ}); for all $\gamma$ the
Jacobian matrix at this point is
\[
\left[
\begin{array}
[c]{cc}%
-2 & 0\\
0 & -2
\end{array}
\right]  ,
\]
therefore $\left(  -\gamma/\sqrt{6},2\right)  $ is a stable improper node (`a
sink star' in other terminology). It is interesting that this node represents
a flat bulk fluid with a $\gamma$ that falls inside the range of acceptable
values as dictated by the energy conditions. It turns out that for all
$\gamma$, this equilibrium is a global attractor of all solutions of
(\ref{nls1})-(\ref{nls2}), see Figure \ref{node1}, where the attractor is
shown for the cases of a cosmological constant ($\gamma=-1$) and a massless scalar field bulk ($\gamma=1$).

\begin{figure}[tbh]
\begin{center}
\includegraphics[scale=0.9]{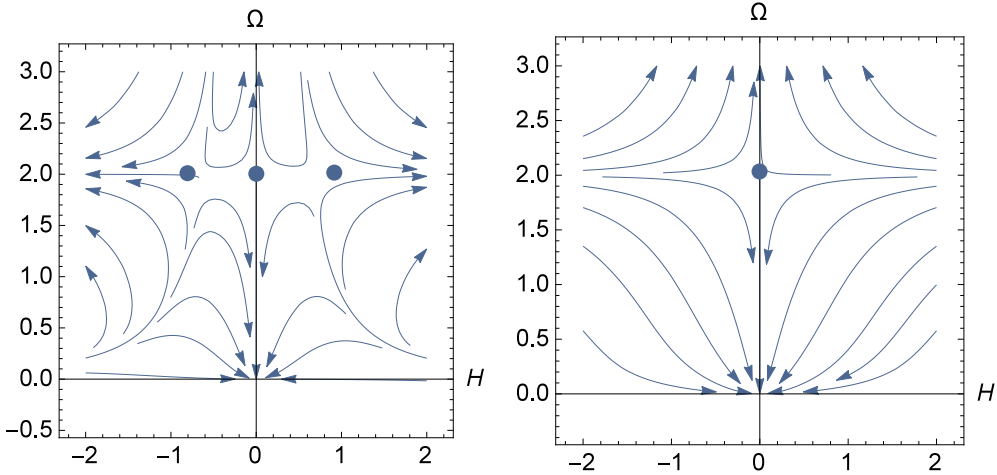}
\end{center}
\vspace*{-10mm}
\caption{Phase portrait of the system (\ref{nls1})-(\ref{nls2}) for
$\lambda=2$ and $\gamma=-1$ and $\gamma=1$. There are three
saddles at $(-1,2)$ and at $(1,2)$ for $\gamma=-1$, but only one at $(0,2)$ for $\gamma=1$.}%
\label{lambda2}%
\end{figure}

We now proceed with the analysis of the case $\lambda\geq1$.
As already mentioned, the system (\ref{nls1})-(\ref{nls2}) has two equilibrium points, located
at the origin and at the phase point $(0,2)$, that is the whole bulk dynamics
is organized around a static empty dS bulk and a static flat bulk fluid. We
note that the linearized system around $(0,0)$ becomes,
\begin{equation}
\left[
\begin{array}
[c]{c}%
H^{\prime}\\
\Omega^{\prime}%
\end{array}
\right]  =\left[
\begin{array}
[c]{cc}%
-1 & 0\\
0 & -2
\end{array}
\right]  \left[
\begin{array}
[c]{c}%
H\\
\Omega
\end{array}
\right]  ,\label{linzed}%
\end{equation}
therefore the point $\left(  0,0\right)  $ is a stable node, and the local
phase portrait of the full nonlinear system (\ref{nls1})-(\ref{nls2}) is
topologically equivalent to that of the linear system (\ref{linzed}). By
inspection of the Jacobian matrix at $\left(  0,2\right)  $ we see that its
eigenvalues are $\pm2$, therefore this equilibrium is a saddle point, that is,
trajectories approaching this point eventually move away. A typical phase
portrait for $\lambda\geq1$ is given in Figure \ref{lambda2} for a
cosmological constant (for $\gamma=-1$) and a massless scalar field bulk (i.e., $p=\rho$).

\begin{figure}[tbh]
\begin{center}
\includegraphics[scale=0.9]{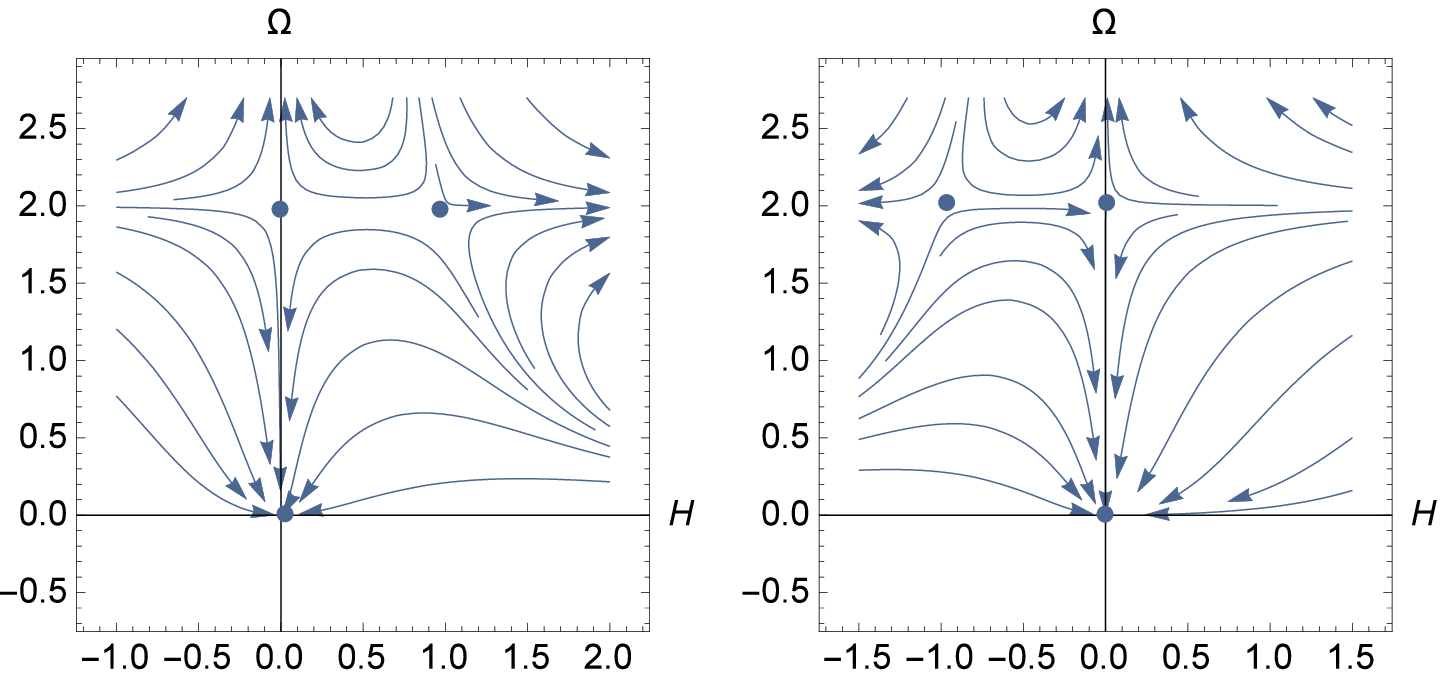}
\end{center}
\vspace*{-10mm}
\caption{Phase portrait of the system (\ref{nls1})-(\ref{nls2}) for
$\lambda=3/2$ and $\gamma=-1/\sqrt{6}$ and $\gamma=1/\sqrt{6}$. In each case there are three
equilibrium points at $\left(  0,0\right)  $, at $\left(
0,2\right)  $ and at $\left(  1/\sqrt{6},2\right)  $ and $\left(  -1/\sqrt
{6},2\right)  $ respectively.}%
\label{saddle1}%
\end{figure}

For $\lambda(n)=1+1/2n$, the equilibria (\ref{equilibria}) lay on
the line $\Omega=2$, at the points where,%
\begin{equation}
H_{\ast}=\left\{  -\frac{1}{\sqrt{6}\gamma},\frac{1}{\sqrt{6}\gamma^{2}%
},-\frac{1}{\sqrt{6}\gamma^{3}},\frac{1}{\sqrt{6}\gamma^{4}},...\right\}
.\label{hequ1}%
\end{equation} We take as typical example the case $\lambda=3/2$. Apart from the
attracting sink at the origin and the saddle at $\left(  0,2\right)  $, the
system has a third equilibrium located according to (\ref{hequ1}) at $\left(
-1/\left(  \sqrt{6}\gamma\right)  ,2\right)  $, and belonging to the first
quadrant for $\gamma<0$, or to the second quadrant for $\gamma>0$. The phase
portrait of the system is shown in Figure \ref{saddle1} for $\gamma
=-1/\sqrt{6}$, $\gamma=1/\sqrt{6}$.

There are three invariant lines, namely $H=0$ and $\Omega=0$ as discussed
after (\ref{nls1})-(\ref{nls2}), as well as the line $\Omega=2$, corresponding to static, empty, and flat bulk fluids respectively. These lines
are the boundaries of the following invariant sets. The strip between the
lines $\Omega=0$ and $\Omega=2$ is an invariant set under the flow of the
system, since every trajectory starting in this strip remains there forever.
Similarly, the sets $\Omega>2,$ $H>0$ and $\Omega>2,$ $H<0$ (that is dynamic AdS bulks) are invariant
sets. For both signs of $\gamma$, all solutions with initial values
$\Omega_{0}>2$ and $H_{0}$ arbitrary, diverge to $\pm\infty$.

The only bounded
solutions observed in Figure \ref{saddle1} are those trajectories approaching the stable node at the origin.
For example, expanding dS scalar field bulks (that is for $\gamma=1$)  with initial values $H_{0}>0$,
$\Omega_{0}<2$ asymptotically approach $\left(0,0\right)$, that is they become static and empty; however, the
determination of the whole basin of attraction of a sink is not always
possible. Finally, there are solutions with $\Omega\left(  \tau\right)  $
approaching the constant value $\Omega_{\ast}=2$ while $H\left(  \tau\right)
$ is diverging to $\pm\infty$, depending on the sign of $\gamma$.

We conclude by giving in the following Table a summary of how the nature of the equilibria given in Eq. (\ref{equilibria}) depends on the type and range of $\gamma$ for the first few values of $n$ in Eq.  (\ref{lambda}):
\begin{center}
\begin{tabular}{ |c|c|c| }
 \hline
 $n$  & $\lambda=1-1/2n$ & $\lambda=1+1/2n$\\
 \hline
 1 & sink for $\gamma\in\left[  -1,1\right]  $ & saddle for $\gamma\in \left[  -1,1\right]  $\\
 2 & sink for $\gamma\in [-1,0)$, saddle for $\gamma\in (0,1]$ & saddle for $\gamma\in \left[  -1,1\right]  $ \\
 3 & sink for $\gamma\in [-1,0)$& saddle for $\gamma\in [-1,0)$\\
 4 & saddle for $\gamma\in \left[  -1,1\right]  $& saddle for $\gamma\in \left[  -1,1\right]  $\\
 5 & sink for $\gamma\in [-1,0)$& saddle for $\gamma\in [-1,0)$\\
 \hline
\end{tabular}
\end{center}
It is interesting to note that because of the presence of a \emph{saddle connection} in Fig. \ref{saddle1} (the horizontal line $\Omega=2$ connecting the two saddles), Peixoto theorem on structural stability is violated.
It also  clearly follows from Figs. \ref{node1}, \ref{lambda2}, \ref{saddle1},  that the three cases, two  corresponding to the  polytropic indices  $\lambda(n)=1\pm 1/2n$, and the third case of $\lambda$ unequal to those, are all qualitatively inequivalent.

We conclude this Section by showing the \emph{impossibility of closed
orbits} for the system (\ref{nls1})-(\ref{nls2}) in the first quadrant of the
$H-\Omega$ plane, that is for expanding, non-empty bulks. 
To see this, we introduce the function,
\begin{equation}
g=\frac{1}{H^a\Omega\,^b},\label{dulac}%
\end{equation}
and the problem is to use  the system (\ref{nls1})-(\ref{nls2}) to determine the constants $a,b$ such that the divergence of the vector field given by the product of the function $g$ times the vector field $(\dot{H},\dot{\Omega})$, that is  $g(\dot{H},\dot{\Omega})$,  is positive for certain ranges of $\lambda,a,b,\gamma$.  The vector field
$g(\dot{H},\dot{\Omega})$ is given by
\begin{equation}
g(\dot{H},\dot{\Omega})=(f_{1}(H,\Omega),f_{2}(H,\Omega)),
\end{equation}
where,
\begin{equation}
f_{1}(H,\Omega)=\frac{1}{H^{a-1}\Omega^b}-\frac{1}{2H^{a-1}\Omega^{b-1}}-3^{\lambda-1}
H^{2\lambda-1-a}\Omega^{\lambda-b},
\end{equation}
and
\begin{equation}
f_{2}(H,\Omega)=\frac{-2}{H^{a}\Omega^{b-1}}+\frac{1}{H^{a}\Omega^{b-2}}+2\gamma3^{\lambda-1}%
H^{2\lambda-2-a}\Omega^{\lambda+1-b}-4\gamma3^{\lambda-1} H^{2\lambda-2-a}\Omega^{\lambda-b}.
\end{equation}
Then, setting $a=1$, the divergence of this vector field is given by,
\be
\begin{aligned}
\operatorname{div}[g(\dot{H},\dot{\Omega})]
=&\frac{2(b-1)}{H\Omega^b}+\frac{2-b}{H\Omega^{b-1}}\\
&+2\gamma 3^{\lambda-1}(2-b)H^{2\lambda-3}\Omega^{\lambda-b}\\
&-4\gamma 3^{\lambda-1} (\lambda-b) H^{2\lambda-3}\Omega^{\lambda-b-1}.
\end{aligned}
\ee
The right-hand-side of this equation is positive provided,
\begin{equation}\label{cond}
\gamma>0,\,\,\,H>0,\,\,\Omega>0,\,\,1<b<2,\,\,\lambda<b,
\end{equation}
so when these inequalities are all true,  the divergence  is strictly positive. The function $g$
from (\ref{dulac}) with this property is  a \emph{Dulac
function}, but no algorithm in general exists for finding such functions. From the inequalities in (\ref{cond}) it then  follows that on the simply connected domain,
\begin{equation}
\mathcal{D}=\{(H,\Omega)|H,\Omega>0\},
\end{equation}
of the planar phase space, the vector field defined by (\ref{nls1})-(\ref{nls2}), satisfies
$(\dot{H},\dot{\Omega})\in\mathcal{C}^{1}(\mathcal{D})$, and $g\in
\mathcal{C}^{1}(\mathcal{D})$, and the divergence $\nabla\cdot g(\dot{H}%
,\dot{\Omega})$ is strictly positive on $\mathcal{D}$. Then by the
Bendixson-Dulac theorem, we conclude that there is no closed orbit lying
entirely on $\mathcal{D}$.


\section{Localisation}

The basic condition for gravity localisation on the brane is that the
4-dimensional Planck mass proportional to the integral $\int_{-\infty}%
^{0}a^{2}dY$ is finite, that is the integral be convergent. We can use
two different approaches to deal with this integral, firstly using the
constraint to re-express the integral in terms of dimensionless variables, and
secondly by direct evaluation.

To start with the first approach, we note that using the Friedmann constraint equation
(\ref{constr}) which is reproduced here,
\[
2-\Omega=\frac{2k}{H^{2}a^{2}},
\]
we can express the `Planck mass integral' $\int a^{2}dY$ in terms of the
dimensionless variables, namely,
\[
2k\int\frac{d\tau}{H^{3}(2-\Omega)}.
\]
One may think that  this integral
expressing the Planck mass can be calculated explicitly, for $H\left(
\tau\right)  $ and $\Omega\left(  \tau\right)  $ given by the solutions
(\ref{omega}) and (\ref{hubble}), however, the constraint equation (\ref{constr}) provides a
relation between the scale factor $a$ and the variables $H$ and $\Omega$ only
for $k\neq0$ models. Therefore we may instead choose to evaluate the integral directly,
\begin{equation}
\int a^{2}dY=\int\frac{a^{2}}{H}d\tau. \label{inte1}%
\end{equation}
Let us consider the case of the linear fluid first. By the definition (\ref{dimless1}), $a\left(  \tau\right)  $ is proportional
to $e^{\tau}$ and $H\left(  \tau\right)  $ is given by the solution
(\ref{hubble}). We are interested to examine whether the Planck mass
integral becomes finite on the interval $(-\infty,0]$. In fact, we are able to
show something more, namely, that it is finite on intervals of the form
$(-\infty,\tau_{1}]$, with a suitable chosen positive $\tau_{1}$ dependent of
the initial conditions and $\gamma$.

It turns out that the integral expressing
the Planck mass can be expressed explicitly in terms of the ordinary
hypergeometric function $_{2}F_{1}\left(  a,b;c;z\right)  $. More precisely,
the value of the indefinite integral (\ref{inte1}) is%
\begin{equation}
\frac{a_{0}^{2}\sqrt{2}e^{3t}\,_{2}F_{1}\left(  a,b;c;z\right)  }{3H_{0}%
\sqrt{2-\Omega_{0}}}, \label{inte2}%
\end{equation}
where%
\be
a=\frac{1}{2},\ \ b=-\frac{3}{4\gamma+2},\ \ c=\frac{4\gamma-1}{4\gamma
+2},\ \ z=\frac{\Omega_{0}\exp\left(  -2\left(  2\gamma+1\right)  \right)
\tau}{\Omega_{0}-2}.
\ee
For some particular values of $\gamma$,  the integral (\ref{inte2}) can be expressed as
combination of elementary functions, although by complicated formulas. We treat the cases $\Omega_{0}\lessgtr 2$ separately below.

For
$\Omega_{0}<2$ the improper integral $\int_{-\infty}^{0}\left(  a^{2}%
/H\right)  d\tau$ exists, i.e. can be expressed in terms of the constants
$a_{0},\Omega_{0},H_{0},$ at least for the representative values,
\be
\gamma=-1,-2/3,-1/2,-1/3,0,1/2,1.
\ee
For the critical value $\gamma=-1/2$, the
hypergeometric function is not defined, but with the solution
(\ref{gammaminus1}), i.e., $H\sim e^{-\tau}$, the integral (\ref{inte1}) is
elementary and $\int_{-\infty}^{0}\left(  a^{2}/H\right)  d\tau=a_{0}%
^{2}/3H_{0}$. (For this value of $\gamma$, the system satisfies all the energy conditions, cf. Ref. \cite{ackr1}, Sect. 5.)

Moreover, in all cases the integral is finite on intervals of
the form $(-\infty,\tau_{1}]$, for some positive $\tau_{1}$. This fact can be
understood, at least for $k\neq0$, if we take into account the remarks after
equation (\ref{hubble}): the integrand $1/H^{3}\left(  \Omega-2\right)  $
remains bounded and approaches zero, even if the functions $H\left(
\tau\right)  $ and $\Omega\left(  \tau\right)  $ take arbitrary large values.

The situation is different if $\Omega_{0}>2$. In that case,  for
$\gamma<-1/2,$ $H\left(  \tau\right)  $ is  real only when the expression inside the root in Eq. (\ref{hubble}) is non-negative, that is when
\be \label{tau*}
\tau\geq\tau_*,
\ee
where,
\be
\tau_*=\frac{1}{-2(2\gamma+1)}\ln\left(\frac{\Omega_0-2}{\Omega_0}\right)<0,
\ee
with equality in (\ref{tau*}) giving the position of the brane. Then we find that  the integral
$\int_{\tau_*}^{\infty}\left(  a^{2}/H\right)  d\tau$ always diverges. For $\gamma\geq-1/2$, $\tau_*$ is positive, but in this case we require  $\tau<\tau_*$ for the expression inside the square root in Eq. (\ref{hubble}) to be positive. Thus  we have to integrate in the range $(-\infty,\tau_*]$, and so we
find that the integral $\int_{-\infty}^{\tau_*}\left(  a^{2}/H\right)  d\tau$
exists, at least for the representative values $\gamma=-1/3,-1/2,0,1/2,1$.

To summarize our results  for the linear fluid: if $\Omega_{0}<2$ the integral (\ref{inte2})
allows for a finite Planck mass for all $\gamma\in\left[  -1,1\right]  $. If
$\Omega_{0}>2$, we have a finite Planck mass for all $\gamma\in\left[
-1/2,1\right]$. In this case,  we choose the upper limit of the integral to
be less than some $\tau_{1}>0$.

\begin{figure}[tbh]
\begin{center}
\includegraphics{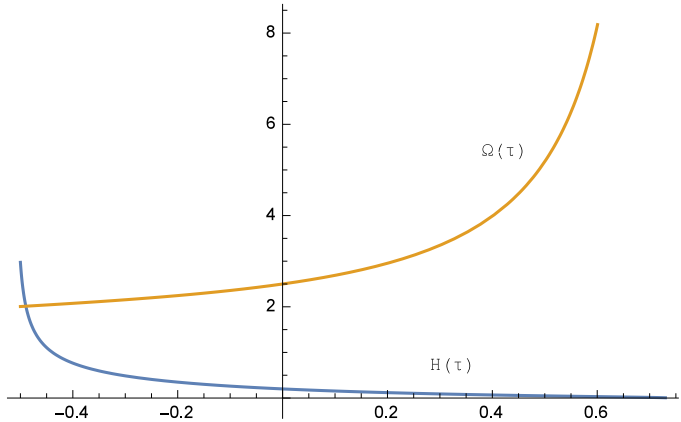}
\end{center}
\caption{Numerical solution for $\gamma=1/2$, $\lambda=2$ and $\Omega_{0}>2$.
$\Omega(\tau)$ develops a singularity at about $\tau\sim0.74$ and $H\left(
\tau\right)  $ has a singularity at $\tau\sim-0.5$.}%
\label{ohnonlinear}%
\end{figure}

Next we consider the case of the nonlinear fluid, $\lambda\neq1$. Equations
(\ref{nls1}) and (\ref{nls2}) can be solved numerically for various values of
the parameters $\gamma$ and $\lambda$ and initial values of the variables $H$
and $\Omega$. In all numerical evaluations the solutions develop finite time
singularities, see for example Figure \ref{ohnonlinear}.

Nevertheless, it
seems that the integral (\ref{inte1}) is finite in any interval between the
singularities. This is due to the fact that the integrand $e^{2\tau}/H\left(
\tau\right)  $ remains bounded and approaches zero, even if the functions
$H\left(  \tau\right)  $ and $\Omega\left(  \tau\right)  $ take arbitrary
large values.  
The numerical investigation described above indicates that even
for the nonlinear fluid, the Planck mass expressed by (\ref{inte1}) may be
finite, although we were unable to prove this result rigorously.

\section{Discussion}
In this paper we have introduced and studied the consequences of a new formulation for the  dynamics of a 4-braneworld embedded in a bulk 5-space. This formulation transforms the problem into a two-dimensional dynamical system that depends on parameters such as the EoS parameter and the degree of nonlinearity of the fluid. This allows us to study the phase space of the model, and also consider in detail the importance of different states, points in phase space, such as the origin or the $(0,2)$-state for the overall dynamical features of the bulk fluid.

For the case of a bulk fluid with a linear equation of state, our new formulation leads to a partial decoupling of the dynamical equations of this model. This in turn implies that the linear fluid case can be solved exactly, and the asymptotic nature of the $(H,\Omega)$ solutions to be directly revealed as well as their dependence on the EoS parameter and the initial conditions to be explicitly shown. In addition, we find that the equilibria of the system depend on the fluid parameter $\gamma$ and this has a major effect of the global dynamics of the system, not present in the simpler case of relativistic cosmologies. The main effect is the existence of a transcritical bifurcation around the $\gamma=-1/2$ value which change the nature of the local equilibria as well as their stability. We also concluded that the overall geometry of the orbits swirls around the two states we call empty bulk and flat fluid, as well as a number of other equilibria.

For the case of a  nonlinear bulk fluid, the dynamics is organized differently for different $\lambda$-values, and shows a preference for polytropic bulk fluids. For instance, the existence on an overall attractor appears only for $\lambda=1/2$, while the dynamics for a bulk having $\lambda>1$ is characterized by portraits organized around nodes and saddle connections for the values of $\lambda=1+ 1/n$. This means that the nonlinear case has a variety of instabilities as well as stable and saddle behaviours. The non-existence of closed orbits in the first quadrant is also a marked feature of the nonlinear bulk fluid.

However, as numerical evaluations  show, despite the existence of singularities, the phenomenon of brane-localization in the sense of having a finite Planck mass is  \emph{self-induced} by the dynamics itself: restricting the dynamics on the orbifold leads generically to gravity localization on the brane. For this conclusion,  although it follows clearly from various numerical evaluation that we have explicitly performed, we have not been able to provide a full analytic proof. We note, however,  that for nonlinear fluids when the null energy condition is satisfied and $\gamma<0$, there are indeed solutions without finite-distance singularities as two of us have shown  in \cite{ack21a} using different techniques such as representation of solutions through hypergeometric expansions and matching.

It would be interesting to extend some of these results further. For example,  to the case of a bulk filled with a self-interacting  scalar field instead of the fluid. Another extension is to study the ambient problem and allow  for singularities at infinity using similar methods as those discussed here. It would also be interesting to study in detail the case $\lambda<1$ where the vector field is non-polynomial. These more general problems will be given elsewhere.

\section*{Acknowledgements}
IA would like to thank the hospitality and financial support of SISSA and ICTP where this work was partially done.

\end{document}